\documentclass{aa}  

\usepackage{graphicx}
\usepackage{lipsum}
\usepackage{xcolor}
\usepackage{natbib}
\usepackage{hyperref}
\usepackage{array}
\usepackage{txfonts}
\newcommand{\rainbow}{{\sc Rainbow\ }}
\newcommand{\lc}{{\tt light\_curve\ }}

\begin{document}

\title{\rainbow: a colorful approach on multi-passband light curve estimation}

   \author{E.~Russeil\inst{1}\thanks{\email{etienne.russeil@clermont.in2p3.fr}},
           K.~L.~Malanchev\inst{2,3,4},
           P.~D.~Aleo\inst{2,5},
           E.~E.~O.~Ishida\inst{1},
           M.~V.~Pruzhinskaya\inst{1},
           E.~Gangler\inst{1},
           A.~D.~Lavrukhina\inst{6},
           A.~A.~Volnova\inst{7},
           A. Voloshina\inst{1},
           T. Semenikhin\inst{3,6},
           S.~Sreejith\inst{8},
           M.~V.~Kornilov\inst{3,9},
           \and
           V.~S.~Korolev\inst{10}          
           (The SNAD team)}

   \authorrunning{Russeil et al.} 
   
   \institute{Universit\'e Clermont Auvergne, CNRS/IN2P3, LPC, F-63000 Clermont-Ferrand, France
        \and
        Department of Astronomy, University of Illinois at Urbana-Champaign, 1002 West Green Street, Urbana, IL 61801, USA
        \and
        Lomonosov Moscow State University, Sternberg Astronomical Institute, Universitetsky pr.~13, Moscow, 119234, Russia
        \and
        McWilliams Center for Cosmology, Department of Physics, Carnegie Mellon University, Pittsburgh, PA 15213, USA
        \and
        Center for Astrophysical Surveys, National Center for Supercomputing Applications, Urbana, IL, 61801, USA
        \and
        Lomonosov Moscow State University, Faculty of Space Research, Leninsky Gori 1 bld. 52, Moscow 119234, Russia
        \and
        Space Research Institute of the Russian Academy of Sciences (IKI), 84/32 Profsoyuznaya Street, Moscow, 117997, Russia
        \and
        Physics Department, Brookhaven National Laboratory, Upton, NY 11973, USA
        \and
        National Research University Higher School of Economics, 21/4 Staraya Basmannaya Ulitsa, Moscow, 105066, Russia
        \and
        Independent researcher
        }

 
  \abstract
   {Time-series generated by repeatedly observing astronomical transients are generally sparse, irregularly sampled, noisy and multi-dimensional (obtained through a set of broad-band filters). In order to fully exploit their scientific potential, it is necessary to use this incomplete information to estimate a continuous light curve behavior. Traditional approaches use \textit{ad-hoc} functional forms to approximate the light curve in each filter independently (hereafter the \textsc{Monochromatic} method).}
   {We present \textsc{Rainbow}, a physically motivated framework which enables simultaneous multi-band light curve fitting. It allows the user to construct a 2-dimensional continuous surface across wavelength and time, even in situations where the number of observations in each filter is significantly limited. 
   }
   {Assuming the electromagnetic radiation emission from the transient can be approximated by a black-body, we combined an expected temperature evolution and a parametric function describing its bolometric light curve. These three ingredients allow the information available in one passband to guide the reconstruction in the others, thus enabling a proper use of multi-survey data. We demonstrate the effectiveness of our method by applying it to simulated data from the Photometric LSST Astronomical Time-series Classification Challenge (PLAsTiCC) as well as real data from the Young Supernova Experiment (YSE DR1).}
   {We evaluate the quality of the estimated light curves according to three different tests: goodness of fit, time of peak prediction and ability to transfer information to machine learning (ML) based classifiers.  Results confirm that \rainbow leads to equivalent (SNII) or up to 75\% better (SN Ibc) goodness of fit when compared to the  \textsc{Monochromatic} approach. Similarly, accuracy when using \rainbow best-fit values as a parameter space in multi-class ML classification improves for all classes in our sample. An efficient implementation of \rainbow has been publicly released as part of the \lc package at \url{https://github.com/light-curve/light-curve-python}.}
   {Our approach enables straight forward light curve estimation for objects with observations in multiple filters and from multiple experiments. It is particularly well suited for situations where light curve sampling is sparse. We demonstrated here its potential for characterizing SN-like events but the same approach can be used for other classes by changing the function describing the light curve behavior and temperature representation. In the context of the upcoming large scale sky surveys and their potential for multi-survey analysis, this represents an important milestone in the path to enable population studies of photometric transients. 
   }


   \maketitle
%

\section{Introduction}
\label{sec:Intro}

Astronomical transients can frequently be described as the observational counterpart of cataclysmic events (e.g. supernovae, hereafter SNe), compact remnant mergers, or the interaction between stars and supermassive black holes (e.g. disruption events, hereafter TDE). Alternatively, they can also be generated by non-periodic bursts in active galactic nuclei \citep{padovani2017}, among other progenitors systems \citep{poudel2022}. They are often initially reported based on their light curves and definitive classification is obtained through spectroscopic follow-up \citep{mura2017, ishida2019b, cannizzaro2021}. 

Detailed study of the evolution of their photometric measurements reveals important aspects of underlying astrophysical properties in individual \citep[e.g.,][]{desai2023}, as well as in population \citep[e.g.,][]{deckers2023} studies. It implies a complete understanding of the objects as 2-dimensional surfaces describing an evolution in time and wavelength. Given caveats intrinsic to astronomical observations, such analysis frequently require some pre-processing stage to extrapolate the surface from the original multi-dimensional, non-homogeneously sampled and noisy light curves, thus enabling detailed subsequent analysis \citep[see e.g.,][]{nyholm2020}. 

This problem is traditionally approach by breaking the 2-dimensional nature of the objects and using an \textit{ad-hoc} parametric function which is fitted to the light curve in each band separately. \citet{zheng2017} used a broken power-law function to estimate the light curve behavior in each filters for SN2011fe; \citet{yan2023} presented an entire software environment for light curve estimation, including many options of parametric fitting as well as semi-analytic models, and \citet{kostyaLC} compared different machine learning based approaches to light curve estimation, with a special emphasis on neural networks.  

A popular parametric function for temporal flux evolution was proposed by \citet{bazin}. It was initially intended to model SNe classes, but it has inspired the development of similar approaches applied to a large variety of objects. A few recent examples include \citet{villar2019}, who proposed an extended parametric function designed to be flexible enough for a broad range of SNe behaviors;
\citet{mccully2022}, who use it to compare properties of  light curves from SN2012Z spanning a decade of observations; 
\citet{muthukrishna2022} employed its resulting goodness of fit as a proxy for anomaly detection; \citet{kelly2023} used it to model and calculate time delays from 5 different light curves from the same strong-lensed SN; \citet{Corsi2023} applied it to model broad-line SN Ic and \citet{fulton2023} used it to obtain a continuous extrapolation for the SN light curve associated with GRB21009A.  

The use of suitable parametric functions have also been applied as a feature extraction procedure for classification tasks. After confronting a given parametric expression with the data, the best-fit parameter values are used as features for machine learning applications. In the case of multi-passband survey, this procedure is repeated independently for every passband in order to completely extract the required information \citep[e.g.,][]{karpenka2013, ishida2019}. Hereafter we name this procedure the \textsc{Monochromatic} method. This approach ensures a full description of the object without any physical assumption, but it comes with 3 important downsides. i) the number of features extracted will linearly scale with the number of passbands considered. In some cases the number of filters can be quiet large, for example, the Vera Rubin Observatory Legacy Survey of Space and Time\footnote{\url{https://www.lsst.org/}}  \citep[LSST,][]{LSST_paper} will use 6 passbands, while the current Southern Photometric Local Universe Survey\footnote{\url{https://www.splus.iag.usp.br/}} (S-PLUS) uses 12. Combining observations from different telescopes can further increase the number of parameters to fit; ii) the feature extraction of an object can only be performed if it contains enough data points in each passband to produce a meaningful fit. If one of them is under-sampled, the final description is compromised and the entire object might be dropped from some analysis. This can significantly affect the size of the sample and create a bias towards brighter, well sampled objects; iii) considering each filter individually will result in very correlated features, since fluxes in different wavelengths are not completely independent. For instance, transient peak times in different passbands are highly correlated. In a machine learning context, having multiple strongly correlated features should be avoided, as it might lead to decrease in overall performances and sparsity in the case of small datasets \citep{mlcorrelation}. 

The issue of independent passband paradigm has been tackled in the literature by using Gaussian process (GP). For example, 
 \citet{boone2019} described SNe in the wavelength versus time parameter space and uses the GP representation for data base augmentation, while \citet{kornilov2023} used it to construct a fully data-driven multi-dimensional representation of superluminous SNe light curves. However, such approaches are non-parametric and computationally expensive, constituting a bottle-neck for high volume data processing.

The arrival of large scale sky surveys, such as LSST, will increase by at least a few orders of magnitude the number of available light curves, boosting the potential impact of such studies but also imposing a new challenge. Ideally, analysis methods should be both: physically motivated and computationally efficient. Thus allowing not only easy application to data releases of increasing volume, but also enabling immediate integration with community broker systems\footnote{\url{https://www.lsst.org/scientists/alert-brokers}}.

One such approach is presented here. The \rainbow method is composed of 3 main ingredients: a black-body profile hypothesis for the emitted electromagnetic radiation; and parametric analytical functions describing its temperature evolution and its bolometric light curve. The coherent combination of these elements results in an efficient description of light curve behavior across wavelengths, even in the presence of sparsely sampled light curves coming from different observational facilities. 
We demonstrate the performance of the \rainbow method when applied to simulated data from the Photometric LSST Astronomical Time-series Classification Challenge \citep[PLAsTiCC,][]{plasticc}\footnote{\url{https://www.kaggle.com/c/PLAsTiCC-2018}} and real data   \citep[spectroscopically confirmed SNe light curves from the Young Supernova Experiment, hereafter YSE][]{Aleo2023}). 

The present work is organized as follows: In Section \ref{sec:Metho} we describe our motivations, experiment design choices (Sect. \ref{sec:bolo} and \ref{sec:Temp}), and fitting procedure (Sect. \ref{sec:fex}). Section \ref{sec:Data} introduces the data set used. In Section \ref{sec:results} we present the results through several metrics, including goodness of fit (Sect. \ref{sec:good}), maximum flux time prediction (Sect. \ref{sec:maxt}) and full or rising light curves classification (Sect. \ref{sec:classif}). We describe results of the method applied to the YSE data in Section \ref{sec:patrick}. Finally, we present our conclusions in Section \ref{sec:conclusion}. Complementary figures are available in Appendix \ref{app:cm} and \ref{app:plot}.

\section{Methodology}
\label{sec:Metho}

We used a black-body spectral model as a proxy for the thermal-electromagnetical behavior of the astrophysical transient.
The observed spectral flux density in this model is characterized by any two of three physical quantities (or their combination): object's solid angle, temperature and bolometric flux.
We chose to parameterize it via two independent functions of time: one for the bolometric flux $F_\mathrm{bol}(t)$ and one for the temperature $T(t)$, which leads to the following expression for the observed spectral flux density per unit frequency $\nu$:
\begin{equation}
F_{\nu}(t, \nu) = \frac{\piup\,B\left(T(t),\nu\right)}{\sigma_\mathrm{SB}\,T(t)^{4}} \times F_\mathrm{bol}(t),
\label{eq:maineq}
\end{equation}
where $B$ is the Planck function and $\sigma_\mathrm{SB}$ is the Stefan--Boltzmann constant. Note that here we do not take into account cosmological effects since the black-body spectrum retains its properties when it is redshifted. Thus, all the quantities here, including temperature, are assumed to be in the observer frame.

For the purpose of comparison with observations, we compute the average of the spectral flux density $F_\nu$ for each passband $p$.
This is done by incorporating its corresponding transmission function, denoted as $R_p(\nu)$:

\begin{equation}\label{eq:flux-passband}
  F_p(t) = \frac{\int F_\nu(t, \nu) / \nu \, R(\nu) \, \mathrm{d}\nu}{\int 1/\nu \, R(\nu) \, \mathrm{d}\nu}\,.
\end{equation}

Notably, instead of integrating over passband transmission, this method could instead be used with the black-body intensity at the effective wavelength of each given passband. Such an approach would be more efficient computationally though less accurate. In this paper we always average over passbands as shown by Eq.~\eqref{eq:flux-passband} using transmissions provided by the SVO Filter Profile Service~\citep{2012ivoa.rept.1015R}.

We name the method \rainbow after its continuous wavelength description (illustrated in Appendix \ref{app:plot}). Equation \ref{eq:maineq} is a general form and can be adapted to different problems by choosing appropriated parametric functions to describe the bolometric flux $F_\mathrm{bol}(t)$ and the temperature evolution $T(t)$. Theses choices determine the type of object described, the number of free parameters and the complexity of the equation. The method is particularly well suited to describe poorly sampled data, as the physical assumptions will coherently fill the lack of information available. It simultaneously addresses the three problems previously highlighted: i) the number of parameters will remain constant independently of the number of passbands considered;
ii) an object correctly sampled overall but with sparse data in one or more passbands can still be fitted. Any information coming from the under sampled filters will still help to constrain the minimization; iii) since a single fit is performed, repetitive information coming from the same parameters across different passbands is avoided. The previous small variability encompassed within the repeated parameters is now contained more densely in the temperature evolution parameters.

It is important to emphasize that, in a high cadence scenario, the physical assumptions can impose certain behaviors that will not exactly correspond to the reality of the data. Thus, \rainbow will not produce optimal results for extremely well sampled light curves. Moreover,  the small number of parameters comes at the cost of adding extra free parameters brought by the temperature evolution function. Therefore this method is only useful in the case of multi-passband data. Nevertheless, \rainbow is perfectly suited to deal with the diverse range of cadences and filter sets adopted by modern data sets.

\subsection{Bolometric flux}
\label{sec:bolo}

\begin{figure}
    \centering
    \includegraphics[width=0.5\textwidth]{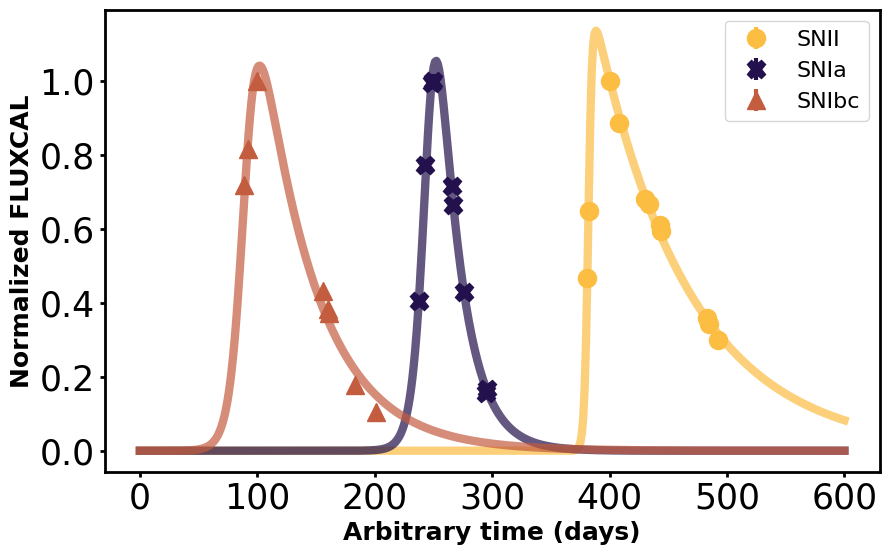}
    \caption{Example of Bazin fits on 3 different types of SNe from the PLAsTiCC dataset, in the LSST-$r$ filter: SNIbc (brown triangles), SNIa (blue crosses) and SNII (yellow circles). For each light curve the flux is normalized by its maximum measured value. The time (horizontal axis) is arbitrarily shifted to display all 3 examples in a single frame. Error bars are included but contained within the points.}
    \label{fig:bazin}
\end{figure}

We used the Bazin functional form \citep{bazin} to represent the bolometric light curve behavior. It describes an exponentially rising and an exponentially decreasing light curve. While not being perfect, especially for plateau phases, it is a reasonably good approximation for different SNe types as illustrated in Figure \ref{fig:bazin}.

We use the following form of the Bazin function:
\begin{equation}
\label{eq:bazin}
f(t) = A \times \frac{\exp{\frac{-(t-t_{0})}{t_\mathrm{fall}}}}{1+\exp{\frac{t-t_{0}}{t_\mathrm{rise}}}}.
\end{equation}
It contains 4 free parameters: $t_\mathrm{rise}$ and $t_\mathrm{fall}$ describe the rising and declining rates; $t_{0}$ is a reference time, which is related to the time of peak,  $t_\mathrm{peak} = t_{0} + t_\mathrm{rise} \times \ln(t_\mathrm{fall}/t_\mathrm{rise}-1)$, and $A$ is an amplitude parameter. We do not include a baseline parameter since the flux baseline is set to zero throughout our analysis.

\subsection{Temperature}
\label{sec:Temp}

\begin{figure}
    \centering
    \includegraphics[width=0.5\textwidth]{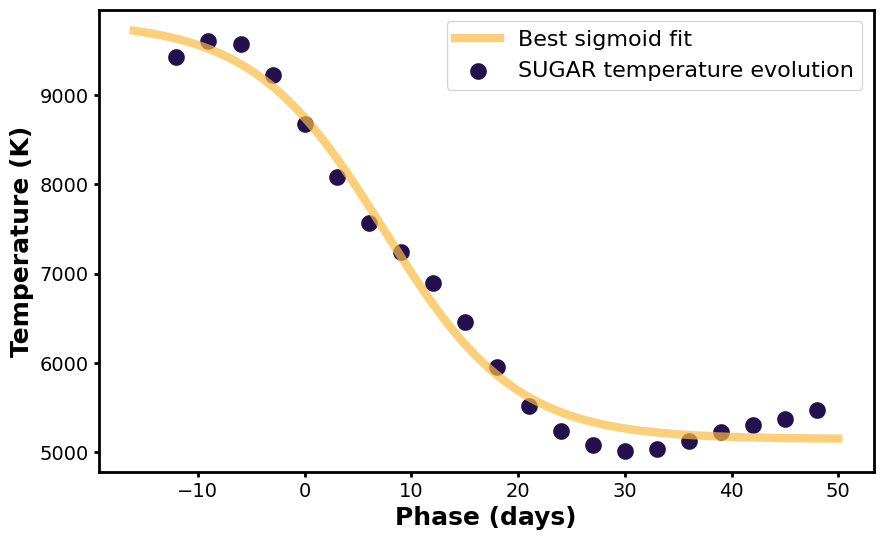}
    \caption{Best logistic function fit on temperature evolution extrapolated from the SUGAR model. Data points have been computed by fitting black bodies to the SNIa spectra from 3250 to 8650 Å at regular time intervals. The phase correspond to the time since maximum brightness.}
    \label{fig:sigmoid}
\end{figure}

 Despite being used to describe different types of transients, the Bazin function is most commonly applied on SNe. Consequently we used a temperature evolution function coherent with SN classes. Using the SUpernova Generator And Reconstructor model (SUGAR, \citealt{Sugar}), a SNIa spectral energy distribution model, we took a spectrum from -12 to 48 days since maximum brightness in B-band with a step of 3 days. Each spectrum was fitted with a black-body function to obtain the temperature at a given time, effectively reconstructing the temperature evolution (Figure \ref{fig:sigmoid}).
We visually inspected the results from a sigmoid-like fit (Equation \ref{eq:temp}) and confirmed it gives a good first approximation for the temperature behavior, while still staying relatively simple and general. Thus, we parameterized the temperature evolution as
\begin{equation}
T(t) = T_\mathrm{min} + \frac{\Delta T}{1 + \exp{\frac{t-t_{0}}{k_\mathrm{sig}}}},
\label{eq:temp} 
\end{equation}

\noindent a four parameter logistic function behaving as two flat curves linked by an exponential slope. $T_\mathrm{min}$ is the minimum temperature that the object will reach, $\Delta T$ describes the full amplitude of temperature, $k_\mathrm{sig}$ corresponds to the slope of the exponential and $t_{0}$ is a reference time parameter which corresponds to the time at half of the slope. Preliminary results using independent $t_{0}$ for Equation \ref{eq:bazin} and \ref{eq:temp} often resulted in almost equally fitted values. Therefore we assume that $t_{0}$ from both equations are equal and merge them into a single parameter.
We use Equation~\ref{eq:temp} to describe the temperature evolution in Equation~\ref{eq:maineq}.

\subsection{Feature extraction}
\label{sec:fex}

As a preprocessing step, each light curve was normalized by the maximum flux measured in LSST-$r$ passband (for the data described in Section \ref{sec:Data}) or ZTF-$r$ (for the data described in Section \ref{sec:patrick}), which ensures a more uniform dataset. Then, we fit the light curves and extract the resulting best-fit parameters. The minimization step is performed using least squares from the \textsc{iminuit} \texttt{Python} library\footnote{\url{https://iminuit.readthedocs.io/en/stable/}} \citep{roos1975}.

When using the \textsc{Monochromatic} approach, we performed a Bazin fit (Equation \ref{eq:bazin}) for each filter separately. Given the functional form, we extract 4 parameters per passband considered. Additionally, we collected the least square loss of the fitting process as a feature for each passband. The $r$-band flux normalization factor previously computed is included as well. Therefore, we extract $n_{features} = 5 n_{passbands} + 1$ features for each object.

In the context of the \rainbow method, we performed a single fit of all the filters at the same time using Equation \ref{eq:maineq}, resulting in 7 best-fit parameters. Additionally we keep the least square loss of the fitting process and the normalizing factor. Therefore, we extract 9 features for each object, independently of the data set considered.

\section{Data}
\label{sec:Data}

We performed benchmark tests on data from the PLAsTiCC dataset \citep{plasticc}. PLAsTiCC was a classification challenge which took place in 2018, and whose goal was to mimic the data situation as expected from LSST. 

Each astronomical source is represented by a  noisy, and non-homogeneously sampled light curve in six different filters\footnote{\url{https://www.lsst.org/about/camera/features}}: $u$, $g$, $r$, $i$, $z$ and $Y$. The complete dataset consists of around 3.5 million light curves, representing 17 classes that encompass both transient and variable objects. In principle, \rainbow could handle any number of passbands, however, increasing the number of filters used in the analysis result in greater chances of objects being insufficiently sampled in at least one passband - which implies being non suited for the \textsc{Monochromatic} method and thus preventing a fair comparison of both paradigms. Therefore we choose to work with  three passbands and discard $u$, $z$ and $Y$ for which the blackbody approximation is the least valid \citep{Pierel_2018}.

We selected every well populated transient types within PLAsTiCC, namely SNIa, SNII, SNIbc, SuperLuminous SuperNovae (SLSN) and TDE. The first three SN types were used for all analysis while TDE and SLSN were only added in the classification tests (Section \ref{sec:classif}) with the goal of increasing the complexity of the task. We required each passband to hold at least 4 photometric true detection points \citep[see][Section 6.3 for a full explanation of true detection]{kessler2019}. This cut is imposed by the \textsc{Monochromatic} method based on the Bazin equation (Equation \ref{eq:bazin}). It is the minimal requirement for the minimization to be reasonable and is therefore applied to both methods. Note that \rainbow minimal requirement (at least 7 true detection points across all passbands) is always true if the previous one is verified. Furthermore we require that the time of peak luminosity (defined as PEAKMJD inside PLAsTiCC metadata) is included within the time span of the light curve.

We used only PLASTiCC data from the Wide Fast Deep (WFD) survey strategy. It represents most of the objects from the simulation and is less well sampled than the Deep Drilling Field (DDF) strategy, making it a perfect test case for sparsely sampled data. We performed feature extraction for objects within the WFD sample and stop once we get 1000 objects of each type, or until we run out of objects. Following this procedure we built a database made of 1000 SNIa, 1000 SNII, 468 SNIbc, 1000 SLSN and 372 TDE. 

Additionally, we produced a second, more challenging database using only the rising part of the light curves. We proceed by removing all points after PEAKMJD, which corresponds either to the time of maximum flux in the $B$ passband in the emitter frame for SNIa, or in bolometric flux for the other transients\footnote{Richard Kessler, private communication \label{foot:rick}}. We maintain the same feature extraction method as previously, including the minimal number of points requirement. This resulted in a smaller database containing 626 SNIa, 532 SNII, 317 Ibc, 616 SLSN and 269 TDE.

\section{Results}
\label{sec:results}

We evaluate the performance of \rainbow through a comparison against the standard \textsc{Monochromatic} method. This analysis comprises both, direct and indirect evaluations. The direct method consists in a measurement of the quality of the light curve reconstruction (Section \ref{sec:good}). However, since the reconstruction quality metric does not necessarily reflect the information loss, we also performed two additional indirect tests: the prediction of the peak time (Section \ref{sec:maxt}) and two classification exercises (Section \ref{sec:classif}). 

\subsection{Quality of fit}
\label{sec:good}

This test was performed independently on all SNIa, SNII and SNIbc previously selected, following the methodology proposed in \cite{kostyaLC}. The light curve time span was divided into 5 day long bins. We randomly selected two bins containing at least one true detection point each, from which all photometric points were removed. It effectively results in two random 5 day long gaps in the light curve. After this step, if enough points remain to fulfill the minimum requirement, they are submitted to both feature extraction methods. The removed points are used to compute the agreement between the estimated light curve and the measured values. We used the normalized Root Mean Squared Error based on observed error (nRMSEo) as a quality metric,

\begin{equation}
\label{eq:rmse}
\text{nRMSEo} = \sqrt{\frac{1}{m} \sum_{i} \left[\frac{(y_{i} - \mu(t_{i}))^{2}}{2\epsilon_{i}^{2}}\right]},
\end{equation}
where $t_{i}$ is the time of measurement, $y_{i}$ is the observed flux, $\epsilon_{i}$ is the observed flux error, $\mu(t_{i})$ is the estimated flux at the time $t_{i}$ and $m$ is the number of observations inside the test sample. 

\begin{figure*}
    \centering
    \includegraphics[width=\textwidth]{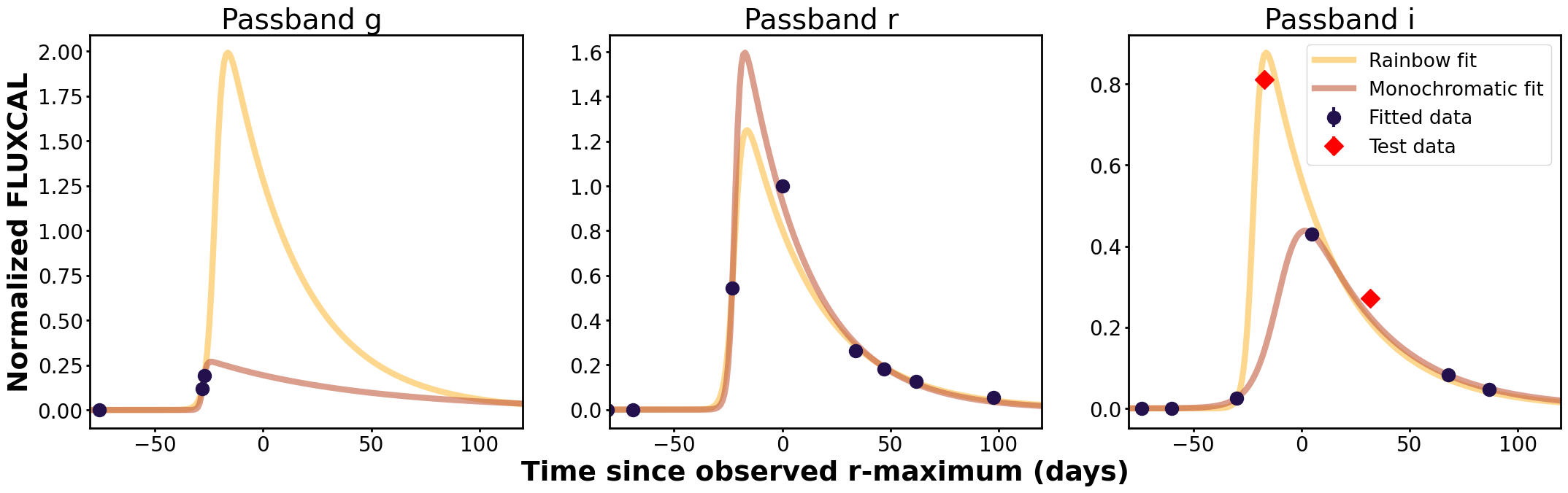}
    \caption{Illustration of the quality of light curve fit on a PLAsTiCC SNIa light curve. Red diamonds represent points  randomly removed and used only to compute the nRMSEo error. The fits were performed considering only points shown as dark blue circles. Error bars are displayed but contained within the points.}
    \label{fig:goodnessoffit}
\end{figure*}

Figure \ref{fig:goodnessoffit} illustrates the procedure and the resulting fits on a given SNIa example for both methods; it highlights how \rainbow is resilient to information loss. One point removed in the $i$-filter carries a lot of information regarding the maximum flux of the object. In $i$-filter, the \textsc{Monochromatic} method misses the peak observation by 20 days with half flux intensity, while \rainbow exploits the information brought by the $r$ passband and produces a more realistic light curve. 

Table \ref{tab:plasticc_goodness} displays the median nRMSEo error for the different classes of SNe. It also provides the 25/75 percentiles. In order to facilitate comparison, we estimate after visual inspection  that an nRMSEo error below 7 can reliably be considered a correct fit for SN-like light curves (the theoretical perfect fit of this metric being $nRMSEo=\frac{1}{\sqrt{2}}$). We observe that overall \rainbow provides a better median error of 4.7, against 5.6 for the \textsc{Monochromatic} method. The improvement of fit quality is clear on SNIa and SNIbc, while both methods perform similarly for SNII.\\
We note that the 25 percentile of errors tend to be higher for the \rainbow method. It means that the very best fits are more often produced by the \textsc{Monochromatic} method. This result is expected since good fits are associated with well sampled data, and we know that the black-body assumption is only a first order approximation. Therefore in a scenario where many data points are available it can act as a constraint on the fit. On the contrary, the 75 percentiles show overall much lower error values of the \rainbow method, which indicates that the method produces less extremely incorrect light curves than the \textsc{Monochromatic} method. This is confirmed by computation of the mean nRMSEo error (less resilient to outliers) for which we obtain 8.8 and 13.1 respectively for the \rainbow and the \textsc{Monochromatic} method.

\begin{table}[]
\begin{center}
\begin{tabular}{ |c||c|c|c|c| } 
\hline
 & SNIa & SNII & SNIbc & All SNe \\ 
\hline
\hline
 \tiny{\textsc{Monochromatic}} & $7.9^{[20.3]}_{[3.3]}$ & $3.7^{[12.2]}_{[1.4]}$ & $4.9^{[12.5]}_{[1.7]}$ & $5.6^{[15.2]}_{[2.1]}$\\
\hline
 \tiny{\rainbow} & $6.8^{[14.3]}_{[3.4]}$ & $3.9^{[8.0]}_{[2.0]}$ & $2.8^{[6.9]}_{[1.5]}$ & $4.7^{[10.3]}_{[2.3]}$\\
\hline
\end{tabular}
\end{center}
\caption{Median nRMSEo (Equation \ref{eq:rmse}) for the \textsc{Monochromatic} method and \rainbow. Lower and upper values respectively represent 25 and 75 percentiles. Results have been computed based on randomly removed bins and are presented for each separate class of SN along with the global result (All SNe).}
\label{tab:plasticc_goodness}
\end{table}

\subsection{Peak time prediction}
\label{sec:maxt}

In this section, we consider a regression task aimed to estimate the time of maximum flux as an indirect approach to evaluate the quality of the light curves reconstructions. PLAsTiCC metadata contains the value PEAKMJD corresponding of the time of the peak flux. It is defined by the time of maximum flux in the $B$ Bessel passband in the source rest frame for SNIa (in that case we use PLAsTiCC redshift metadata), or in received bolometric flux for every other transient\footref{foot:rick}. We predict this value using a direct and an indirect method.\\

\begin{figure}
    \centering
    \includegraphics[width=0.5\textwidth]{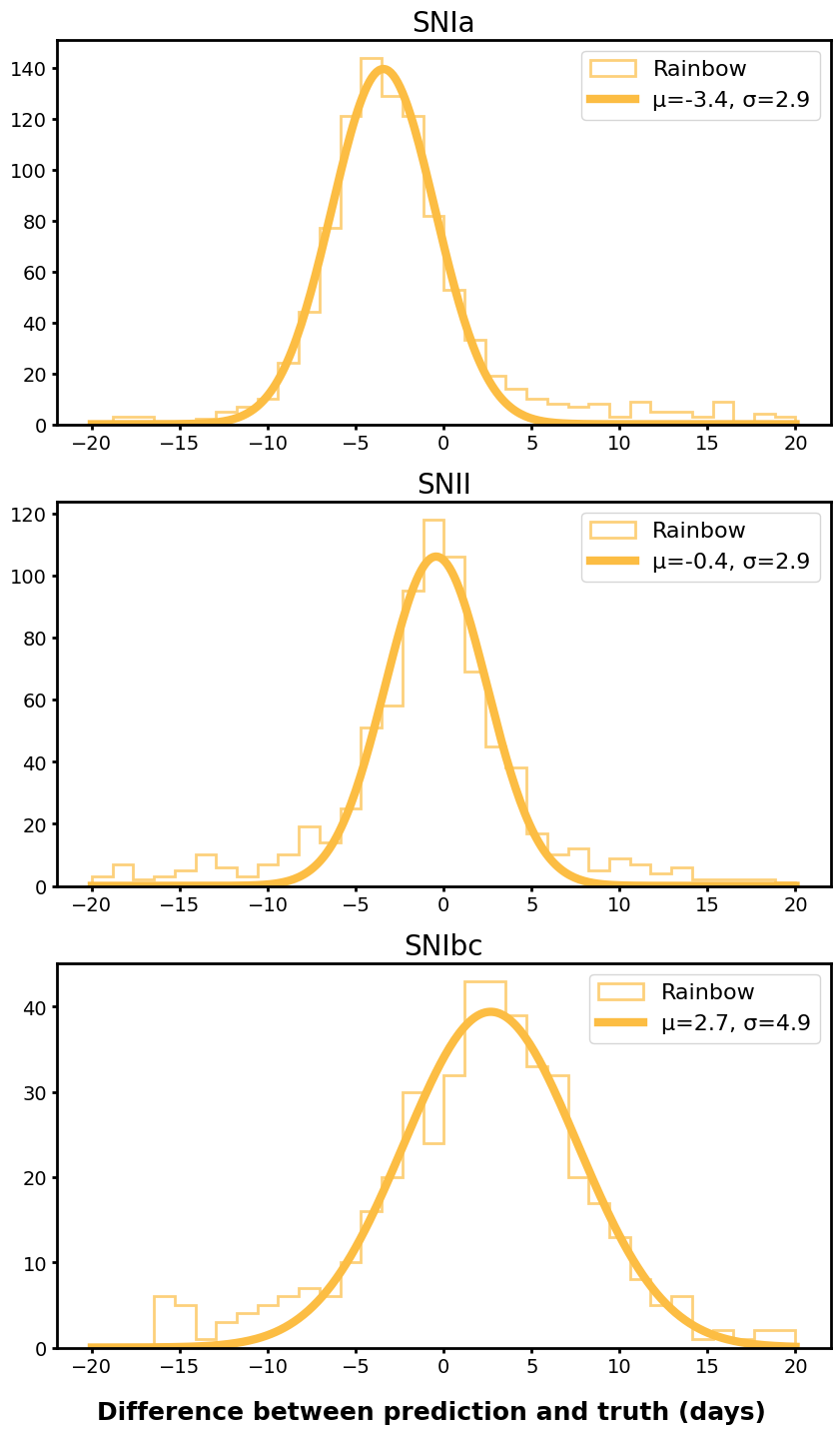}
    \caption{Distributions per class of SNe of the difference between \rainbow prediction of time of maximum and reported time of maximum as given within the PLAsTiCC metadata. \rainbow prediction is directly computed from the light curve estimation using the definition of PLAsTiCC PEAKMJD. Additionally a Gaussian is fitted to the distributions to evaluate its mean and standard deviation.}
    \label{fig:tmax_direct}
\end{figure}

The direct prediction method highlights the versatility of \textsc{Rainbow}. The two dimensional fit function (Equation \ref{eq:maineq}) gives access to the flux in any passband at any time. Additionally \rainbow provides a direct estimation of the bolometric flux of the object. Therefore, in the case of SNIa, we can directly compute the time of maximum flux in the $B$ passband even if no measurements were taken at this wavelength. For the other transients, we can use the bolometric flux explicitly encoded in the equation. In this analysis we predicted the PEAKMJD for SNIa, SNII and SNIbc. Figure \ref{fig:tmax_direct} presents the distribution in the difference between the prediction and the reported peak time per class. We fit the distribution with a Gaussian to obtain a mean and a standard deviation. We observe that the method provides unbiased results for SNII. SNIa and SNIbc predictions are biased respectively around 3 days before and after the peak. While the bias exist, an error of 3 days constitute a small error when it comes to peak prediction of the considered SNe classes. Overall the results show a small spread especially for SNIa and SNII with a standard deviation of 2.9 days.

\begin{figure}
    \centering
    \includegraphics[width=0.5\textwidth]{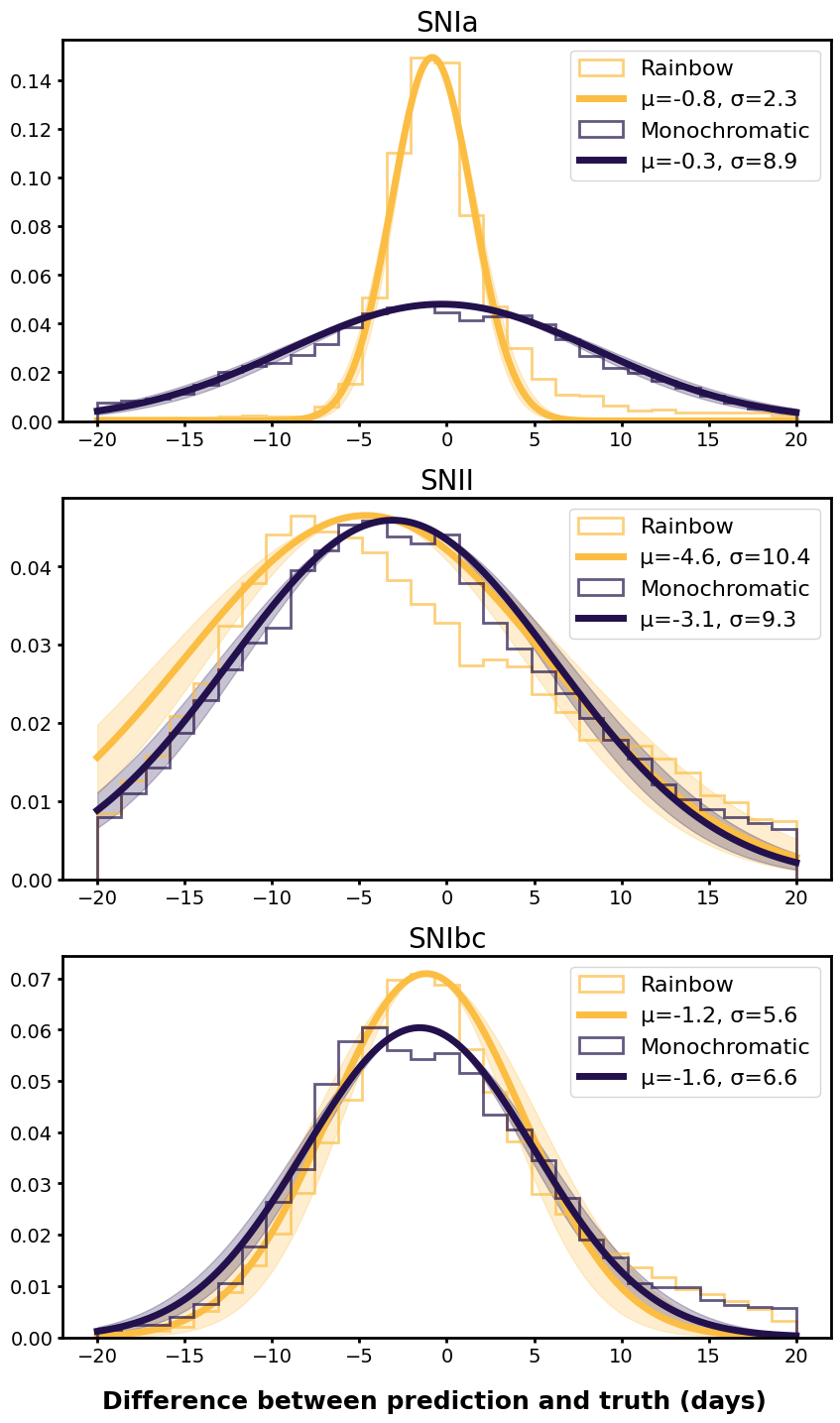}
    \caption{Distributions per class of SNe of the difference between prediction of time of maximum and true time of maximum as given within the PLAsTiCC metadata. Predictions have been computed based on linear regressor models trained with \rainbow (yellow)  and \textsc{Monochromatic} (dark blue) features. Additionally a Gaussian is fitted to the distributions to evaluate its mean and standard deviation.}
    \label{fig:tmax}
\end{figure}

This method has the advantage of offering a direct computation of the time of maximum, however, it can only be computed for the \rainbow method. The \textsc{Monochromatic} feature extraction procedure does not provide information about the bolometric nor the $B$ passband flux. In order to fairly compare the quality of prediction, we compute a second indirect metric using the scikit-learn linear regression\footnote{\url{https://scikit-learn.org/stable/modules/generated/sklearn.linear_model.LinearRegression.html}} algorithm trained on the features extracted from both methods. We leave all hyper parameters on default values to ensure a fair comparison between the two. 
We evaluate the performances on 100 iterations  \citep{Modele_selection} of bootstrapping \citep{bootstrapping}, which is sufficient to produce robust results. Figure \ref{fig:tmax} displays histograms of all bootstrapping predictions combined for each SN class. Additionally we fit a Gaussian to the distribution result of each bootstrapping iteration from which we extract the standard deviation and the mean. Gaussians presented in Figure \ref{fig:tmax} represent the mean Gaussian of the bootstrapping iterations. The colored area around the Gaussians represent a 1-$\sigma$ deviation of the bootstrapping standard deviation distribution.

Results show excellent predictive power of the \rainbow features for SNIa with a mean of 0.8 and a standard deviation of 2.3 days (Figure \ref{fig:tmax}, top panel). The \textsc{Monochromatic} features resulted in very wide distribution with a standard deviation of 8.9 days. In the case of SNII, both sets of features result in very spreaded predictions. For this particular type using a direct prediction provides a much better indicator on the time of maximum flux. For SNIbc we observe very similar results between both methods, with a little improvement in the standard deviation for the linear regression trained with the \rainbow features. \\

\subsection{Classification}
\label{sec:classif}

A common way to use features extracted from light curves is in machine learning classification tasks \citep[e.g. ][]{Xu2023, Alerce}. An informative set of features provides a good summary description of an object and can be used to distinguish several types of classes. We perform a multi-class classification exercise as a way to measure the relative information quality of the features extracted using the \rainbow method, compared to the those resulting from the \textsc{Monochromatic} procedure. Therefore, our goal is not to create the best possible classifier, but simply to provide a fair comparison on a given dataset. For the classification algorithm, we use the Random Forest implementation from the \texttt{Sklearn} library \footnote{\url{https://scikit-learn.org/stable/modules/generated/sklearn.ensemble.RandomForestClassifier.html}} \citep{scikit-learn}, leaving all hyper parameters unchanged. We decided to perform two classification challenges, using full light curves and using only the rising part.\\

For the first exercise, we sub-sampled the database from Section \ref{sec:Data} and built a balanced dataset of 300 light curves of each class (SNIa,  SNII, SNIb, SLSN and TDE), thus ensuring a uniform representation of the different light curve shapes. We evaluate the results with 100 iterations of bootstrapping. 

\begin{figure}
    \centering
    \includegraphics[width=0.5\textwidth]{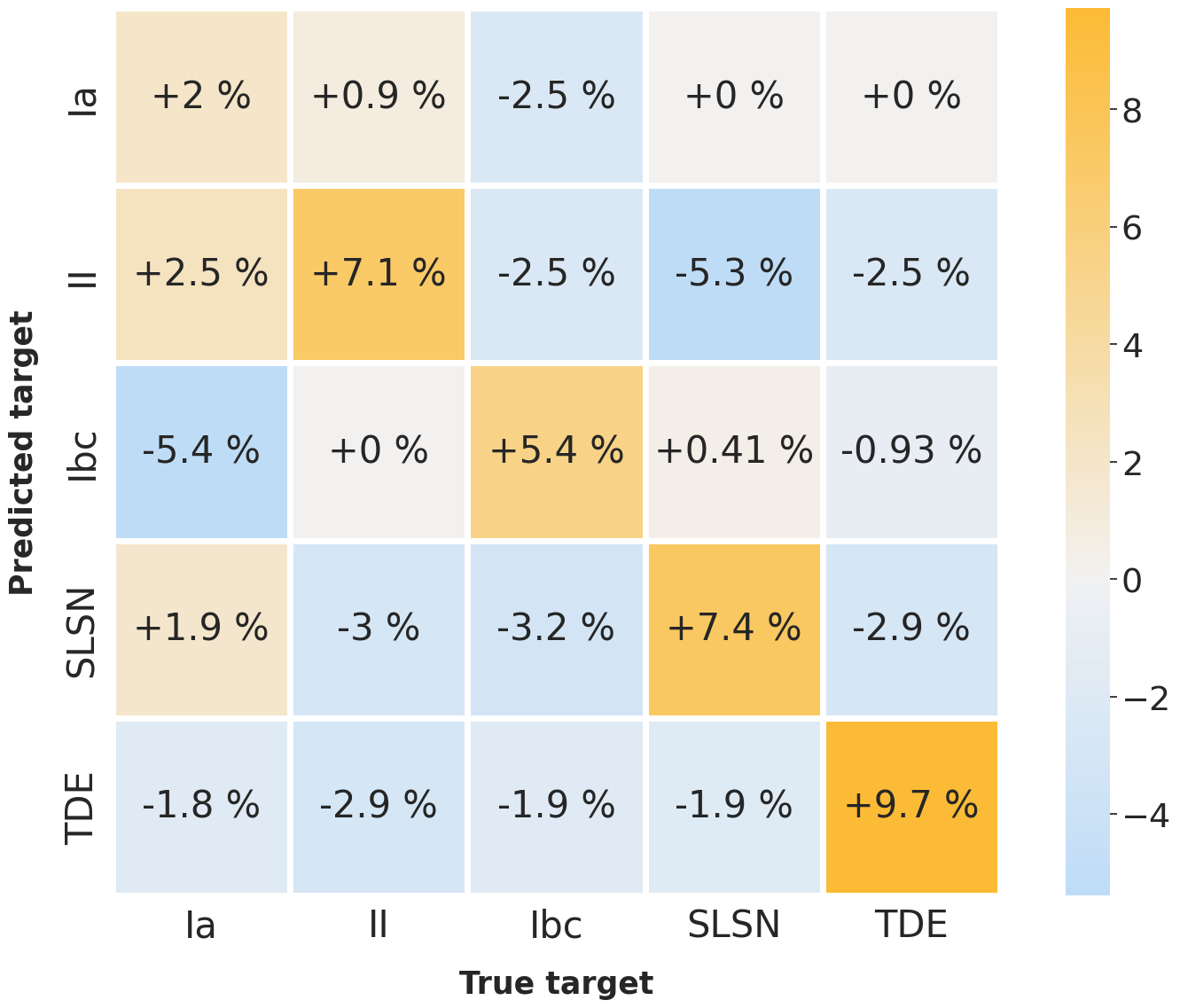}
    \caption{Confusion matrix difference between the Random Forest classifiers trained on \rainbow  and \textsc{Monochromatic} features. The dataset is composed of 300 light curves of each class (SNIa, SNII, SNIbc, SLSN and TDE). Numbers represent the difference (\rainbow - \textsc{Monochromatic}) in the median score of 100 iterations of bootstrapping. Individual confusion matrices for each method are given in Appendix \ref{app:cm} (Figures \ref{fig:absolute_confusion_multi} and \ref{fig:absolute_confusion_multi_bazin}).}
    \label{fig:confusion_multi}
\end{figure}

Figure \ref{fig:confusion_multi} displays the differences in median confusion matrix between the \rainbow features and the standard features. The overall median accuracy values are 88.4\% and 81.9\%, respectively (See Appendix \ref{app:cm},  Figures \ref{fig:absolute_confusion_multi} and \ref{fig:absolute_confusion_multi_bazin} for the individual confusion matrices). Results demonstrate clearly that features extracted with the \rainbow method enclose more discriminative information. Not only the overall accuracy is better, but every single type of transient is more accurately classified. We note that the long lasting transients:  SNII, SLSN and TDE, display the largest improvement when compared to the \textsc{Monochromatic} method. 

\begin{figure}
    \centering
    \includegraphics[width=0.5\textwidth]{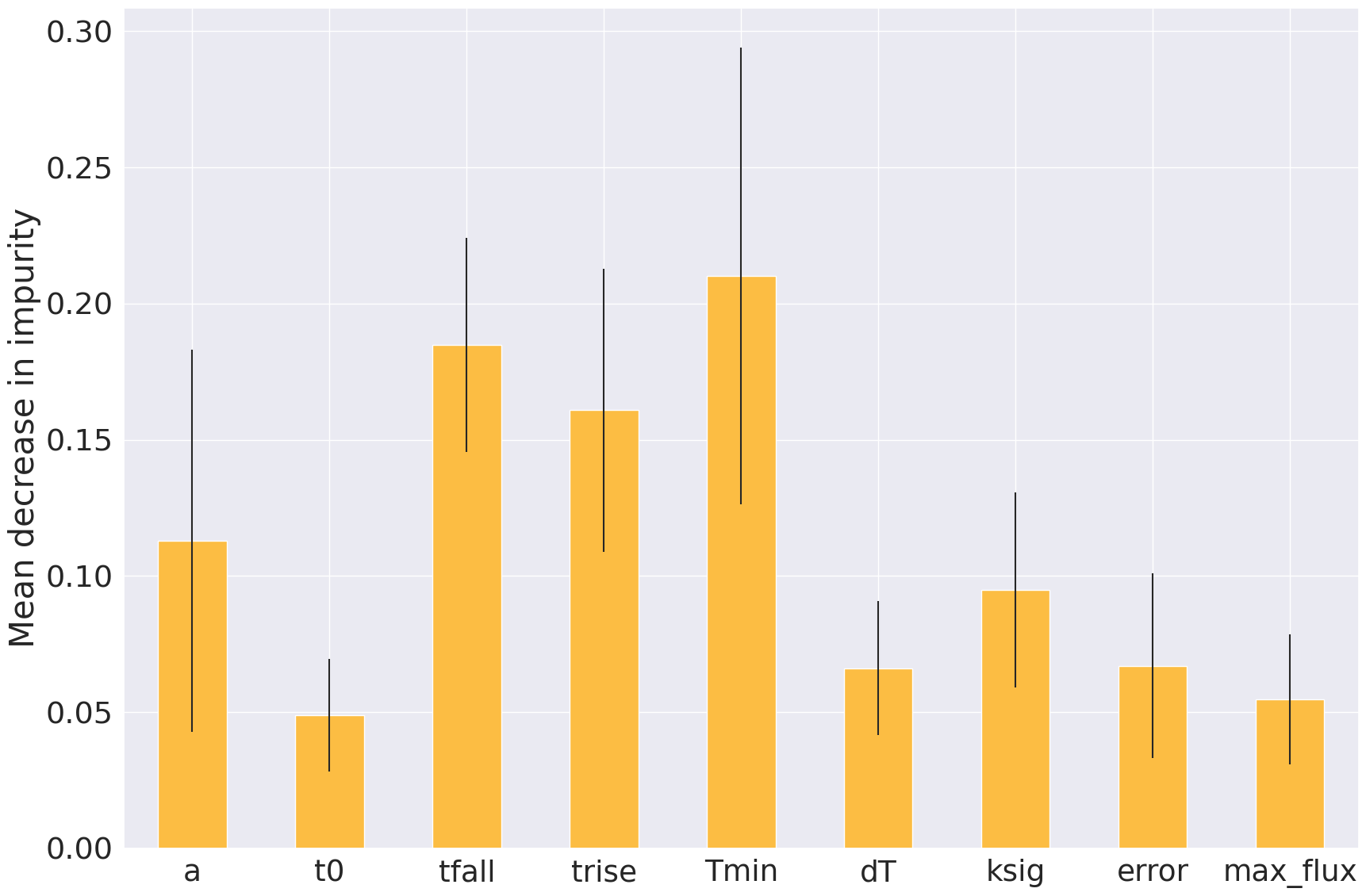}
    \caption{Mean feature importance over 100 bootstrapping iterations of the Random Forest classifier trained with \rainbow features.}
    \label{fig:feature_importance}
\end{figure}

Figure \ref{fig:feature_importance} presents the importance of each \rainbow feature in the classification process. The second and third most important features to separate the different classes are $t_{\rm fall}$ and $t_{\rm rise}$. They are coming from the bolometric function and are predictably core in the description of a transient type. The most relevant feature is $T_{\rm min}$, which describes the minimum temperature that the object will reach after the event. This underlines the importance of the blackbody approximation within the \rainbow model. 

In order to quantify the difference of results between the two classifiers, we perform a corrected McNemar test as proposed in \cite{mlcorrelation}. From this test, we compute a $\chi^2$ that gives the degree of certainty to which we can reject the null hypotheses, i.e that result differences are due to random chance and that both classifiers perform equally well.

\begin{table}
\begin{center}
\begin{tabular}{ |c||c |c| } 
\hline
 & \tiny{Correct \rainbow} & \tiny{Incorrect \rainbow} \\ 
\hline
\hline
 \tiny{Correct \textsc{Monochromatic}}& 416 & 35  \\ 
\hline
 \tiny{Incorrect \textsc{Monochromatic}} & 71 & 29  \\ 
\hline
\end{tabular}
\caption{Mean contingency table of objects correctly and incorrectly classified by the Random Forest models trained with \rainbow and standard features over 100 bootstrapping iterations.}
\label{tab:mc_nemar}
\end{center}
\end{table}

We computed a 2x2 matrix (Table \ref{tab:mc_nemar}) that compares both model predictions to each other (which should not be mistaken to a classical confusion matrix result from a given model). This matrix is computed at each bootstrapping step and the table displays the average over the 100 iterations. The meaningful squares are the top right (TR) and bottom left (BL) since they displayed all the cases where the classifiers outputed different answers. From this we compute the $\chi^2$ metric as:

\begin{equation}
\chi^{2} = \frac{(|TR - BL| - 1)^{2}}{TR + BL}.
\label{eq:mc_nemar} 
\end{equation}

After computation we obtain a $\chi^2=13$, which in the one degree of freedom case results in a p-value of $3\times10^{-4}$. This result shows with more than 3 $\sigma$ confidence that the results differences are not due to random chance. \\ \\

Previous classification results display good performances of the \rainbow method on complete light curves, but in some cases one might want to perform an early classification on a still rising transient in order to make a decision about telescope follow up \citep[e.g.][]{Leoni2022}. Thus, we evaluate the \rainbow method by repeating the classification exercise in this scenario. From the rising light curve database described in Section \ref{sec:Data}, we randomly built a balanced dataset of 250 of each class  (SNIa,  SNII, SNIb, SLSN and TDE). The number of object per class must be reduced to 250 to maintain a balanced dataset. As previously, we applied 100 iterations of bootstrapping, and used a Random Forest algorithm to build a classifier. Similarly to Section \ref{sec:classif}, we computed the differences in median confusion matrix between the \rainbow and the \textsc{Monochromatic} features, as shown in Figure \ref{fig:confusion_multi_rising}. The overall median accuracies are respectively 59.5\% and 50.9\% (See Appendix \ref{app:cm}, Figures \ref{fig:absolute_confusion_multi_rising} and \ref{fig:absolute_confusion_multi_rising_bazin} for the individual confusion matrices). Once again results show that \rainbow features lead to a better classification of each type of rising transients. TDE display the largest difference with 19\% more accuracy. With the exception of SNII, all other rising SNe are clearly better disentangled using the rainbow features. The type SNII includes SNII-P events that are characterised by a slow decreasing plateau phase. In the case of rising light curves we lose this determinant information, which could explain the decrease in the accuracy differences when compared to full  SNII light curve case. 

\begin{figure}
    \centering
    \includegraphics[width=0.5\textwidth]{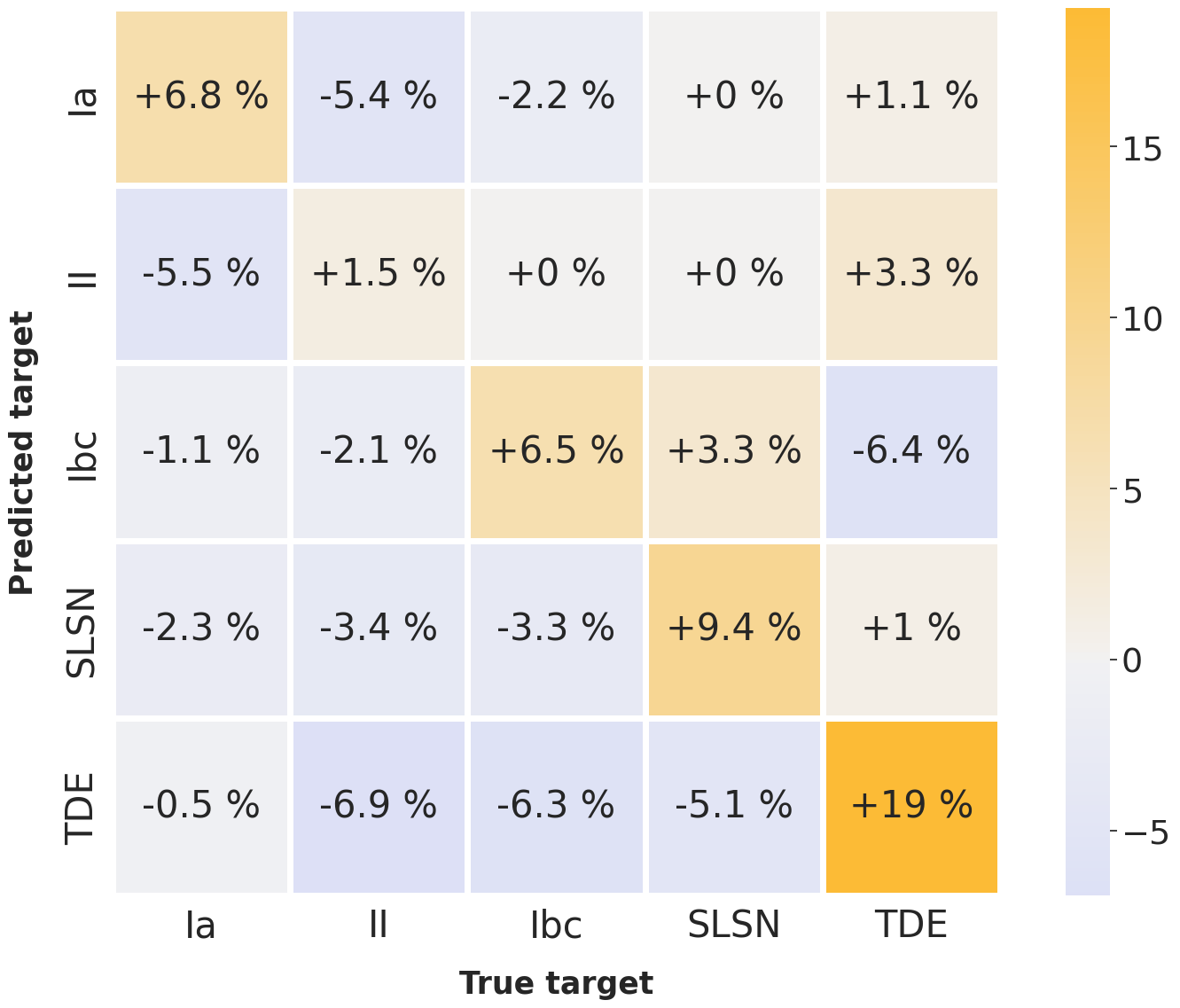}
    \caption{Confusion matrix difference between the Random Forest classifiers trained on \rainbow features and \textsc{Monochromatic} features. The dataset is composed of 250 \textbf{rising} light curves of each class (SNIa, SNII, SNIbc, SLSN and TDE). Numbers represent the difference (\rainbow - Standard) in the median score of 100 iterations of bootstrapping. Individual confusion matrices for each method are given in Appendix \ref{app:cm} (Figures \ref{fig:absolute_confusion_multi_rising} and \ref{fig:absolute_confusion_multi_rising_bazin}).}
    \label{fig:confusion_multi_rising}
\end{figure}

\begin{table}[]
\begin{center}
\begin{tabular}{ |c||c |c| } 
\hline
 & \tiny{Correct \rainbow} & \tiny{Incorrect \rainbow} \\ 
\hline
\hline
 \tiny{Correct \textsc{Monochromatic}} & 213 & 65  \\ 
\hline
\tiny{Incorrect \textsc{Monochromatic}} & 113 & 148  \\ 
\hline
\end{tabular}
\label{tab:mc_nemar_rising}
\caption{Mean contingency table of rising objects correctly and incorrectly classified by the Random Forest models trained with \rainbow and standard features over 100 bootstrapping iteration.}
\end{center}
\end{table}

We also performed a McNemar test for this scenario. The average correct and incorrect classification of both methods are shown in Table \ref{tab:mc_nemar_rising}. Applying Equation \ref{eq:mc_nemar} we obtained a  $\chi^2 = 10.7$, which is equivalent to a p value of $1\times10^{-3}$. The result is again statistically robust and lead to a confidence of more than 3 sigma that our results are not due to random chance. 

\section{Real data application}
\label{sec:patrick}

In this section we demonstrate the capabilities of the method when applied to  a real data multi-survey scenario, and compare it to the \textsc{Monochromatic} procedure. 
We used the Young Supernova Experiment Data Release 1 \citep[YSE DR1, ][]{Aleo2023}, which contains final photometry 
of 1975 transients observed by the Zwicky Transient Facility \citep[ZTF,][]{bellm2019} $gr$ and Pan-STARRS1 \citep[PS1,][]{chambers2016} $griz$. YSE DR1 is the largest available low-redshift homogeneous multi-band dataset of SNe, 
and thus provides the perfect testing ground with a variety of real objects. It encompasses SNe observations across vast timescales, magnitudes, and redshifts: those which last a few days to over a year, ranging in apparent magnitudes $m=[12^m, 22^m]$, and span a redshift distribution up to $z \approx 0.5$. \\

While adding crucial deep photometry, PS1 global cadence is well below that of ZTF, which makes independent filter fit impossible in many cases due to insufficient number of photometric points. This implies that a traditional feature extraction step will result in a significant loss of objects. This issue can be addressed by considering PS1-$gr$ and ZTF-$gr$ as equivalent, effectively stacking the light curves. While this approximation is generally accurate, small discrepancies can be observed because of the differences in passband transmission profiles\footnote{\url{http://svo2.cab.inta-csic.es/svo/theory/fps3/index.php?mode=browse&gname=Palomar&gname2=ZTF&asttype=}}$^{,}$ \footnote{\url{http://svo2.cab.inta-csic.es/svo/theory/fps3/index.php?mode=browse&gname=PAN-STARRS&asttype=}} and difference in photometric pipelines. Therefore, the rigorous approach is to consider passbands independently. In this context and in the context of multi-survey analysis in general, \rainbow offers the possibility to perform fits on all data available at once, without any passband approximation. \newline
We compare \rainbow and the standard method using the evaluation of the quality of the fit described in Section \ref{sec:good}. 
We use all spectroscopically confirmed SNIa, SNII and SNIbc, including non-detections, for a total 254 SNe, available in Zenodo\footnote{\url{https://zenodo.org/record/7317476}} \citep{Aleo2023Zenodo}. Additionally we manually add two non-detection points in ZTF-$gr$ passbands at -300 and -400 days before the measured maximum flux time. It provides a flux baseline that helps constraining the fit, especially in a scenario where only the falling part of the light curve is available. We show results for two cases: one where the dataset was entirely used to perform a \rainbow fit (7 parameters per object); and another where we combined PS1-$gr$ and ZTF-$gr$ light curves are fitted independently with the \textsc{Monochromatic} method (total of 8 parameters per object). Since PS1-$i$ is poorly sampled (with ~30\% of the objects containing less than 4 detection points) and there is no ZTF-$i$ band available, we decide to restrict our test to $gr$ passbands only.\\

\begin{figure*}
    \centering
    \includegraphics[width=\textwidth]{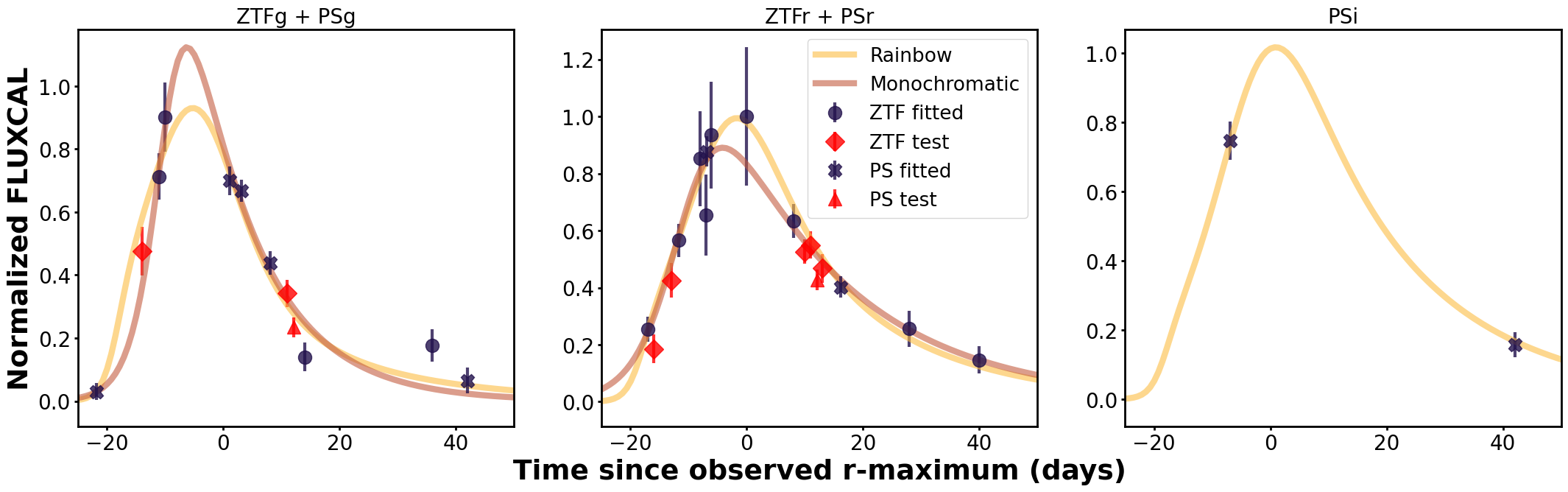}
    \caption{Illustration of the quality of fit evaluation process on a YSE DR1 SNII light curve (SN 2020thx). Red points are randomly removed and used to compute the nRMSEo error. The fits are performed considering the dark points only.}
    \label{fig:goodnessoffit_YSE}
\end{figure*}

\begin{figure}
    \centering
    \includegraphics[width=0.5\textwidth]{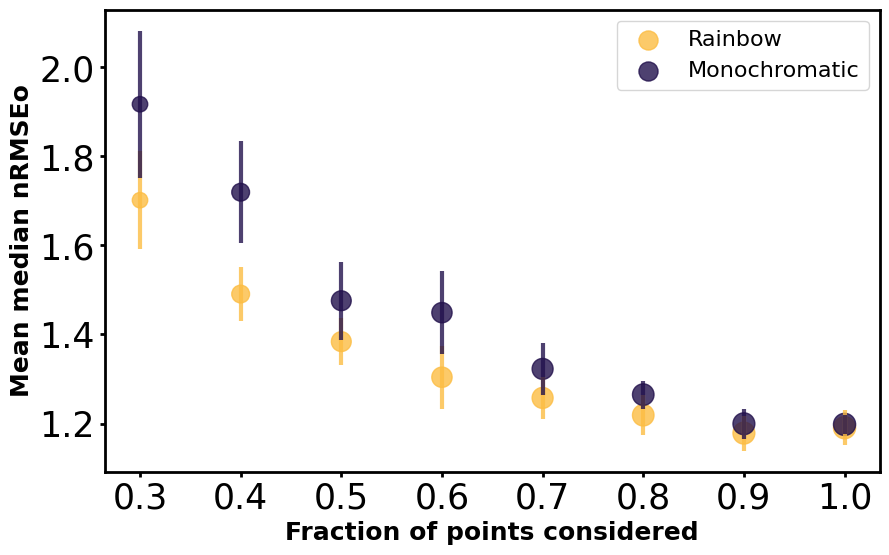}
    \caption{Evolution of the median nRMSEo error for different sizes of sub sample of the YSE dataset. Error bars represent one standard deviation of the 10 random sub samples. Data point surface are proportional to the mean number of SNe in the samples, from 123, for 30\%, to 254, for 100\%.}
    \label{fig:nRMSEo_evol}
\end{figure}

This dataset provides well sampled light curves with on average 15 and 19 detection points respectively on $g$ and $r$ passbands from ZTF and PS1 combined. Since \rainbow is particularly efficient at compensating the lack of information from poorly sampled passbands, we expect both methods to perform equally well in this scenario. Therefore we compared the methods for different sampling levels. We created 8 datasets by randomly sampling from 30\% to 100\% of the points, in steps of 10\%. In order to provide uncertainties we performed  this procedure 10 times with different seeds. Since we still require a minimum number of points per passband (see Section \ref{sec:Data}), the datasets contains on average from 123 SNe, for 30\% sampling, to 254 SNe, for the complete sample. Figure \ref{fig:goodnessoffit_YSE} illustrates the fits on a SNII (SN 2020thx) sampled at 50\%. We display the average median nRMSEo error (Equation \ref{eq:rmse}) per sampling level in Figure \ref{fig:nRMSEo_evol}. The error bars correspond to one standard deviation of the median error over the 10 seeds. The area of each point is proportional to the mean number of SNe in each sample. We observe a clear superiority of \rainbow, especially in extreme situations where the number of points is severely limited. Both methods are equivalent in term of prediction of light curve profile when using 90\% of the original number of points or more. This result indicates that the small additional information provided by PS-$i$ together with the physical assumptions of the \rainbow framework make a crucial difference on sparse transient datasets.

\section{Conclusion}
\label{sec:conclusion}

Estimating a continuous light curve behavior from sparse and noisy photometric measurements is of crucial importance to the study of astronomical transients. When applied to large data sets, it can help understanding the general behavior of underrepresented classes, as well as extreme examples of the most well study ones \citep[e.g.,][]{sanders2015}. Moreover, with the advent of large scale sky surveys like LSST, features from photometric light curves are used to feed machine learning classifiers, since they will be the sole information available for the majority of the observed transients \citep{plasticc}. 

The \rainbow framework presented here produces a continuous 2-dimensional surface in time and wavelength (Appendix \ref{fig:YSE_3D}) even in cases where the light curves are sparsely sampled in a number of different passbands. It starts assuming the thermal-electromagnetic behavior of the transient can be approximated by a black-body, and incorporates user-defined parametric functions representing the temperature evolution and bolometric light curve behavior. As a result, it uses information in the available bands to inform the reconstruction in other wavelengths, providing a simple and robust framework for estimation and feature extraction, which is especially suited for situations where the data is sparse or spread across a set of passbands. 

We used simulated data from PLAsTiCC to demonstrate the effectiveness of the method in a series of tests: goodness of fit (Section \ref{sec:good}), estimation of the time of peak brightness (Section \ref{sec:maxt}) and using the best-fit parameter values as input to machine learning-based classifications (Section \ref{sec:classif}). In all these, we compared \rainbow to the more traditional approach of fitting a parametric function independently to each passband (named here the \textsc{Monochromatic} method). Results show that \rainbow outperforms or equals the results obtained from the traditional method, with the advantage of being applicable to significantly more sparse light curves.

Nevertheless, the method also inherits the drawbacks of the black-body assumption, which may not be suitable for specific science cases. For example the energy distribution of a supernova is far from being a black-body in the ultraviolet and infrared spectrum (e.g., \citealt{faran2018}). The \rainbow method should be used in all wavelengths and epochs for which the black-body assumption holds. Caution should also be applied when dealing with very well sampled light curves. Even if the black-body hypotheses is valid, it remains an approximation to physically describe the gaps in the data. In the case of high cadence measurements, it acts as a constraint that can erase important information. Whenever a large number of data points is available, allowing the necessary number of parameters to be fitted by independently fitting a parametric function to each passband will result in a better approximation. We demonstrated this issue using real data from YSE DR1 (Section \ref{sec:patrick}). Results show that \rainbow provide significantly more informative reconstructions in the presence of small number of observations. 

In the tests presented here, we used a particular set of functions to represent the temperature and bolometric light curve behaviors, however, these can be easily replaced by more suitable ones depending on the class of transients being analysed\footnote{A completely data driven option for the bolometric light curve behavior based on symbolic regression is currently under development and will be the focus of a follow-up work (Russeil \textit{et al.}, in prep).}.  

\rainbow has been incorporated into a well established feature extraction package\footnote{\url{https://github.com/light-curve/light-curve-python}} \citep{malanchev2021} which is already used by three different community brokers: ANTARES \citep{antares}, AMPEL \citep{ampel} and Fink \citep{fink}. Thus it can be immediately used by the community to analyse the alert data.

In the context of modern astronomical surveys, not only the volume of data will pose an important challenge. The quality and complexity of the data gathered by new surveys will also evolve. LSST will push the limits of detection to even fainter magnitudes in 6 different passbands, but the majority of the survey strategy (wide-fast-deep) will probably consist of significantly sparse light curves for at least a few of the passbands \citep{lochner2018}. In this context, \rainbow represents an efficient option to enable modelling and analysis of multi-dimension light curves.

\section*{Acknowledgements}
This work significantly benefited from 2 SNAD workshops: SNAD-V\footnote{\url{https://snad.space/2022/}}, held in Clermont Ferrand - France, in 2022, and SNAD-VI\footnote{\url{https://snad.space/2023/}}, held in Antalya - Turkey, in 2023.
The reported study was funded by RFBR  according to the research project №21-52-15024.  The authors acknowledge the support by the Interdisciplinary Scientific and Educational School of Moscow University “Fundamental and Applied Space Research”. P.D.A. is supported by the Center for Astrophysical Surveys (CAPS) at the National Center for Supercomputing Applications (NCSA) as an Illinois Survey Science Graduate Fellow.

\bibliography{ref}

\appendix

\section{Surface plot}
\label{app:plot}

\begin{figure}
    \centering
    \includegraphics[width=\columnwidth]{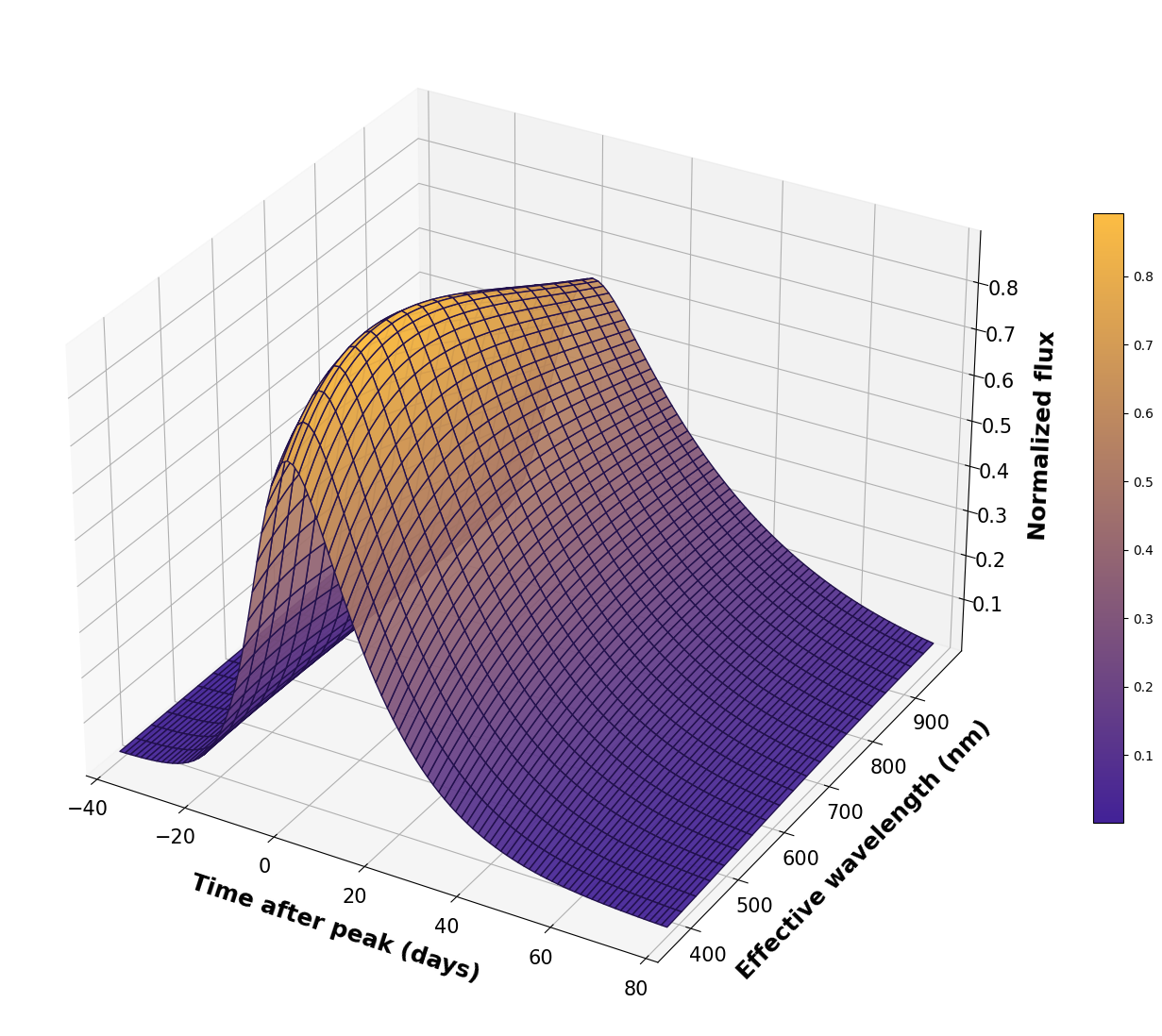}
    \caption{Surface plot representation of a \rainbow fit of a SNII (SN 2020thx) light curve. Representations per passband of the same object with the data points are displayed in Figure \ref{fig:goodnessoffit_YSE}.}
    \label{fig:YSE_3D}
\end{figure}

Figure \ref{fig:YSE_3D} illustrate the shape of a 2-dimensional surface built from real light curves from SN2020thx using the \rainbow approach.

\section{Rainbow classification confusion matrices}
\label{app:cm}

\begin{figure}
    \centering
    \includegraphics[width=0.5\textwidth]{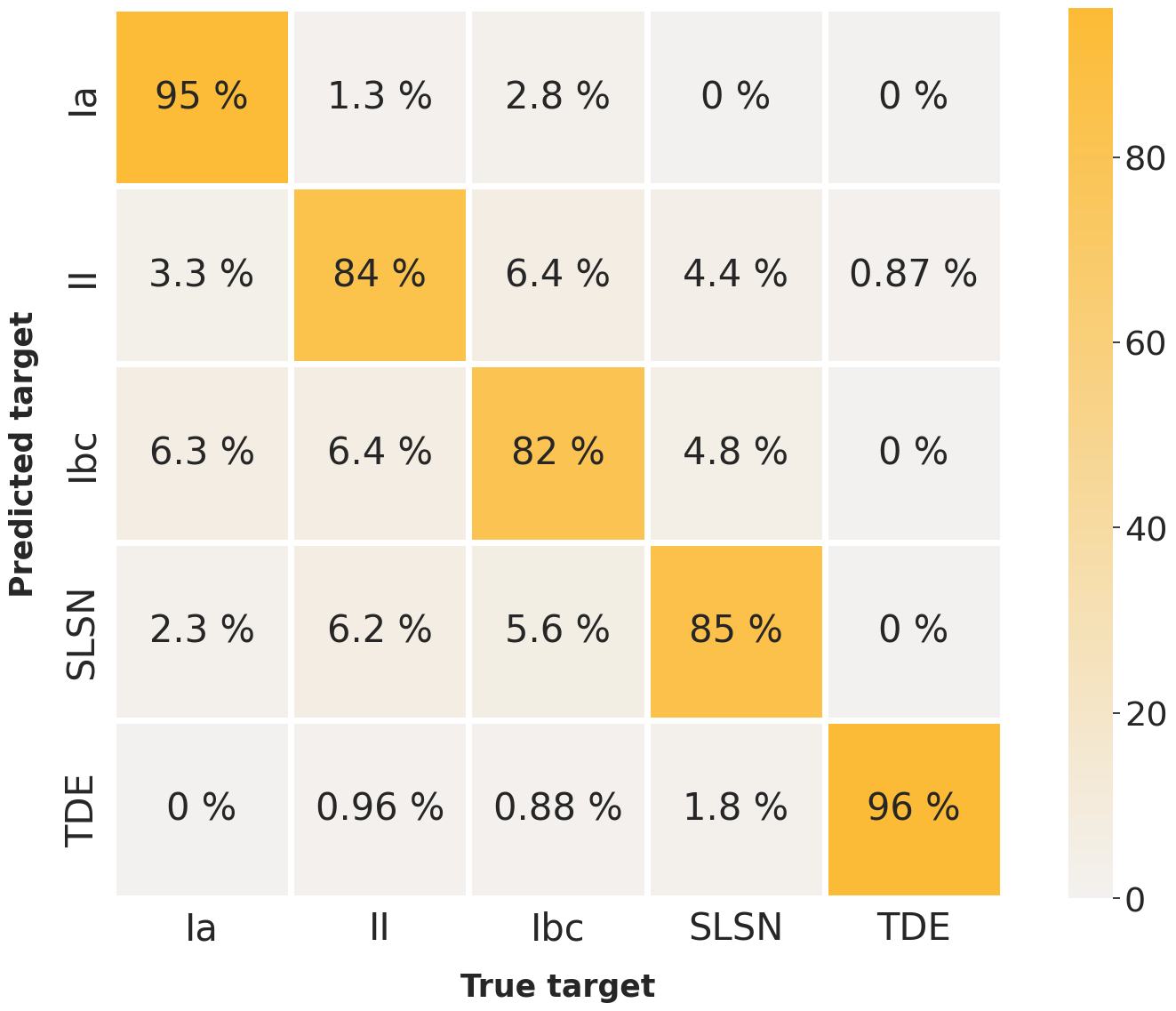}
    \caption{Full light curve scenario. Confusion matrix of the Random Forest classifier trained on \rainbow features. The dataset is composed of 300 light curves of each class (SNIa, SNII, SNIbc, SLSN and TDE). Numbers represent the median score of 100 iterations of bootstrapping. These results were used to construct those shown in Figure \ref{fig:confusion_multi}.}
    \label{fig:absolute_confusion_multi}
\end{figure}

\begin{figure}
    \centering
    \includegraphics[width=0.5\textwidth]{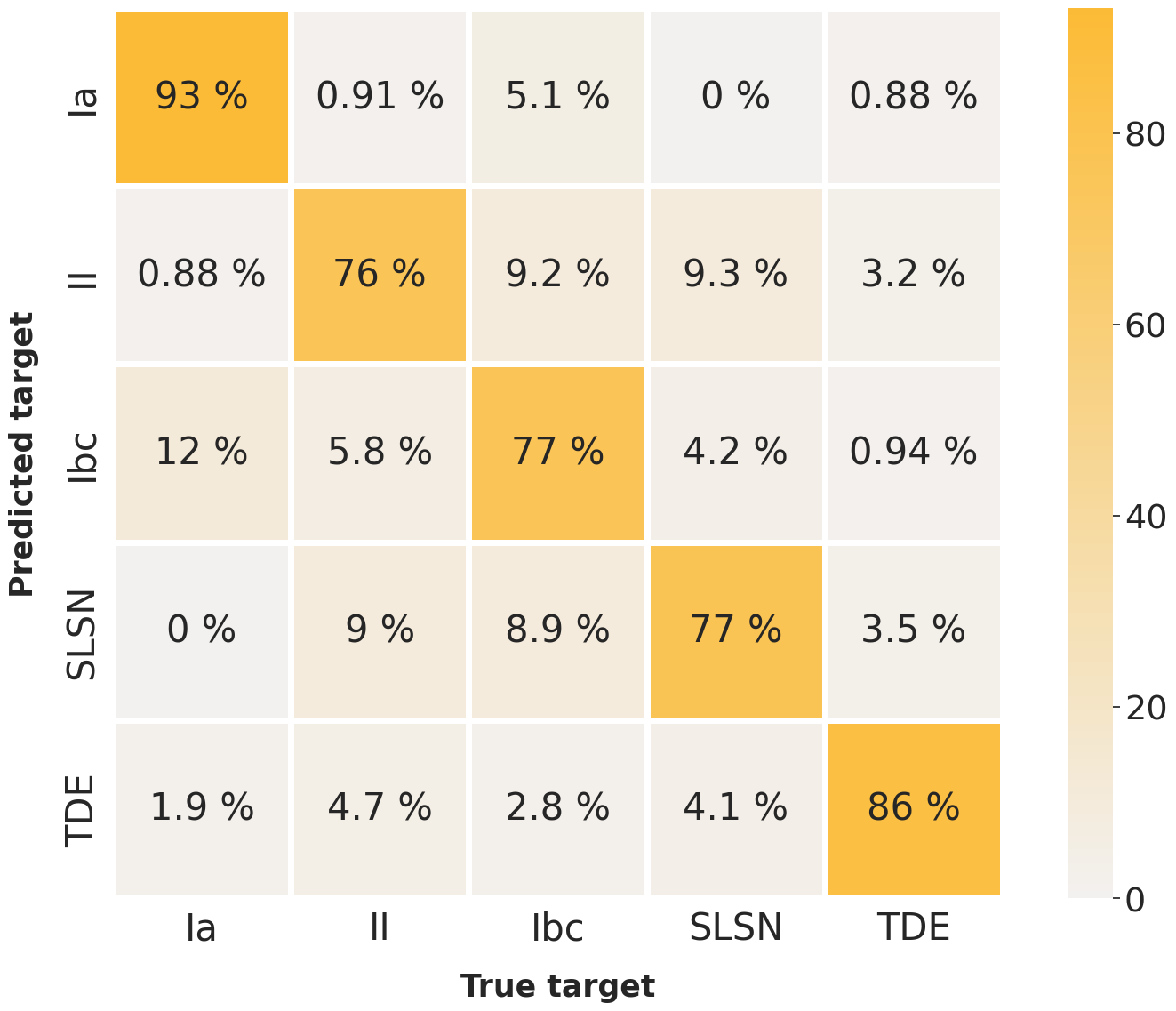}
    \caption{Full light curve scenario. Confusion matrix of the Random Forest classifier trained on \textsc{Monochromatic} features. The dataset is composed of 300 light curves of each class (SNIa, SNII, SNIbc, SLSN and TDE). Numbers represent the median score of 100 iterations of bootstrapping. these results were used to construct those shown in Figure \ref{fig:confusion_multi}.}
    \label{fig:absolute_confusion_multi_bazin}
\end{figure}

\begin{figure}
    \centering
    \includegraphics[width=0.5\textwidth]{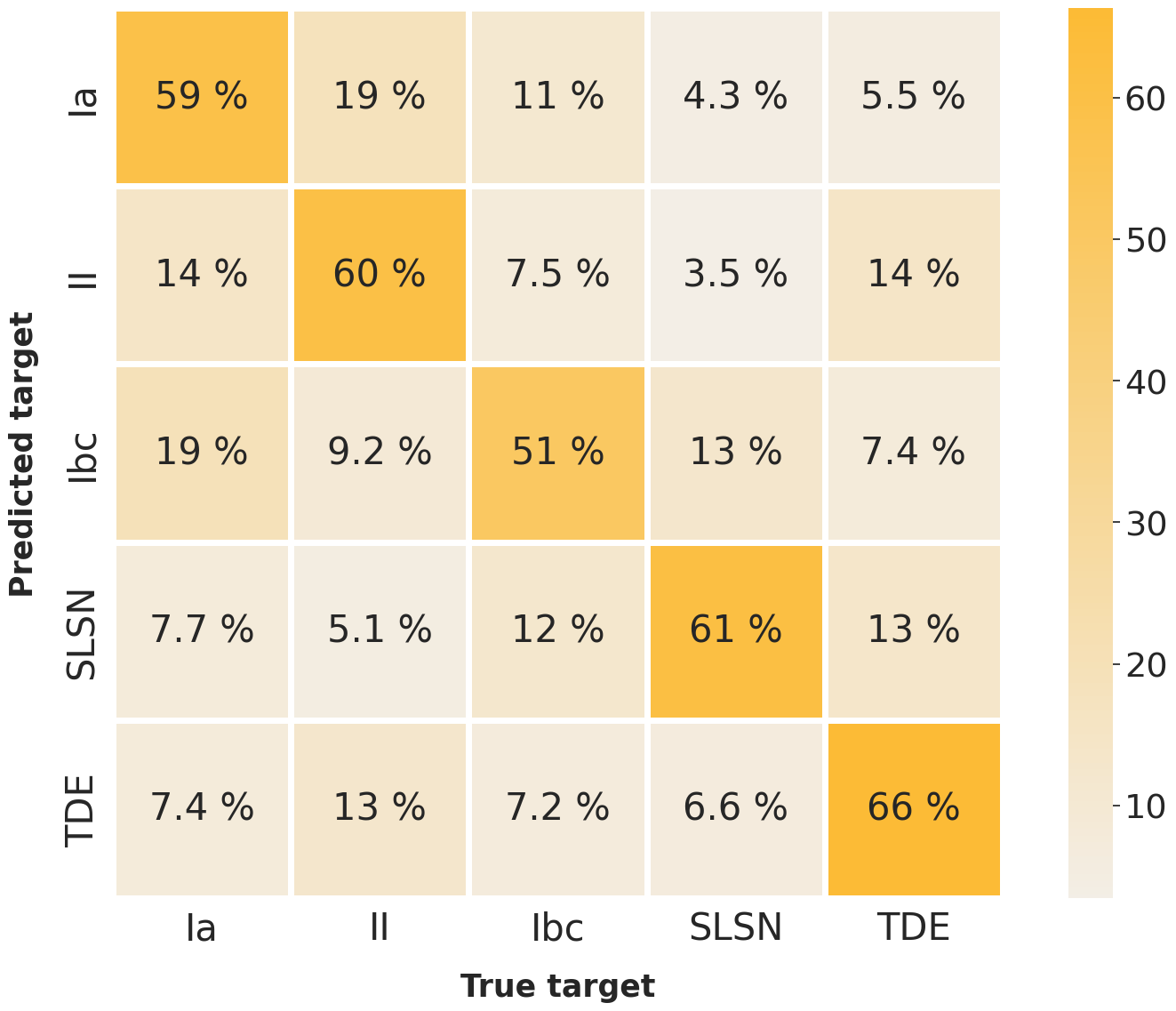}
    \caption{Rising light curve scenario. Confusion matrix of the Random Forest classifier trained on \rainbow features. The dataset is composed of 250 light curves of each class (SNIa, SNII, SNIbc, SLSN and TDE). Numbers represent the median score of 100 iterations of bootstrapping. These results were used to construct those shown in Figure \ref{fig:confusion_multi_rising}.}
    \label{fig:absolute_confusion_multi_rising}
\end{figure}

\begin{figure}
    \centering
    \includegraphics[width=0.5\textwidth]{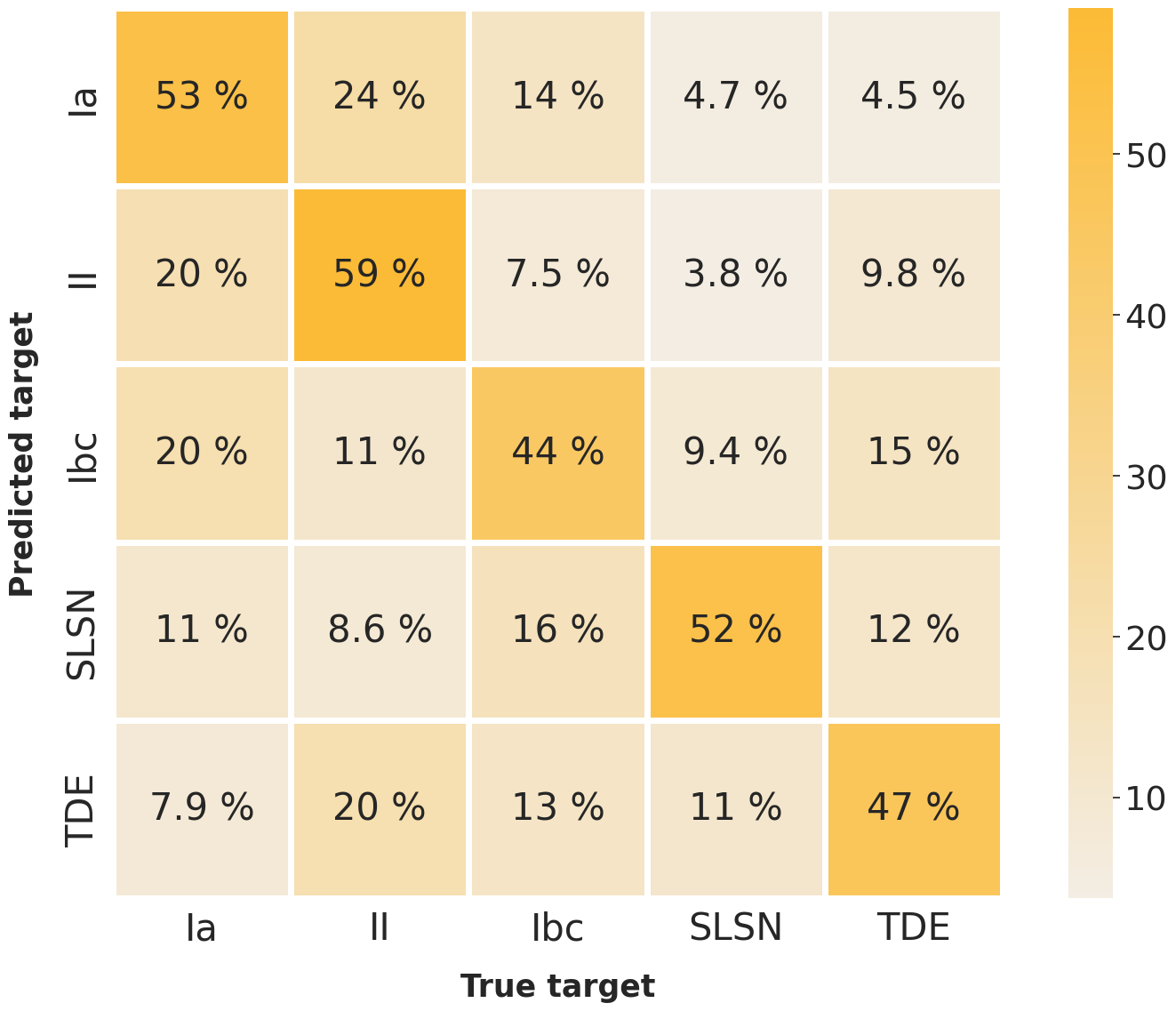}
    \caption{Rising light curve scenario. Confusion matrix of the Random Forest classifier trained on \textsc{Monochromatic} features. The dataset is composed of 250  light curves of each class (SNIa, SNII, SNIbc, SLSN and TDE). Numbers represent the median score of 100 iterations of bootstrapping. These results were used to construct those shown in Figure \ref{fig:confusion_multi_rising}.}
\label{fig:absolute_confusion_multi_rising_bazin}
\end{figure}

Classification confusion matrix of Random Forest models trained on \rainbow (Figures \ref{fig:absolute_confusion_multi} and \ref{fig:absolute_confusion_multi_rising}) and \textsc{Monochromatic} features (Figures \ref{fig:absolute_confusion_multi_bazin} and \ref{fig:absolute_confusion_multi_rising_bazin}). 

\end{document}